\PassOptionsToPackage{unicode}{hyperref}
\PassOptionsToPackage{hyphens}{url}
\documentclass[
  conference]{IEEEtran}
\usepackage{xcolor}
\usepackage{amsmath,amssymb}
\usepackage{iftex}
\ifPDFTeX
  \usepackage[T1]{fontenc}
  \usepackage[utf8]{inputenc}
  \usepackage{textcomp} % provide euro and other symbols
\else % if luatex or xetex
  \usepackage{unicode-math} % this also loads fontspec
  \defaultfontfeatures{Scale=MatchLowercase}
  \defaultfontfeatures[\rmfamily]{Ligatures=TeX,Scale=1}
\fi
\usepackage{lmodern}
\ifPDFTeX\else
\fi
\IfFileExists{upquote.sty}{\usepackage{upquote}}{}
\IfFileExists{microtype.sty}{% use microtype if available
  \usepackage[]{microtype}
  \UseMicrotypeSet[protrusion]{basicmath} % disable protrusion for tt fonts
}{}
\makeatletter
\@ifundefined{KOMAClassName}{% if non-KOMA class
  \IfFileExists{parskip.sty}{%
    \usepackage{parskip}
  }{% else
    \setlength{\parindent}{0pt}
    \setlength{\parskip}{6pt plus 2pt minus 1pt}}
}{% if KOMA class
  \KOMAoptions{parskip=half}}
\makeatother
\usepackage{longtable,booktabs,array}
\usepackage{calc} % for calculating minipage widths
\usepackage{etoolbox}
\makeatletter
\patchcmd\longtable{\par}{\if@noskipsec\mbox{}\fi\par}{}{}
\makeatother
\IfFileExists{footnotehyper.sty}{\usepackage{footnotehyper}}{\usepackage{footnote}}
\makesavenoteenv{longtable}
\providecommand{\tightlist}{%
  \setlength{\itemsep}{0pt}\setlength{\parskip}{0pt}}
\usepackage[numbers]{natbib}
\usepackage{tikz}
\usetikzlibrary{arrows.meta,positioning,fit,backgrounds,calc}
\usepackage{booktabs}
\usepackage{array}
\usepackage{amsmath}
\usepackage{amssymb}
\usepackage{fontspec}
\usepackage{fancyvrb}
\usepackage{fvextra}
\fvset{fontsize=\scriptsize,breaklines=true,breakanywhere=true}
\let\oldverbatim\verbatim
\renewcommand{\verbatim}{\scriptsize\oldverbatim}
\usepackage{newunicodechar}
\newunicodechar{ψ}{$\psi$}\newunicodechar{ω}{$\omega$}
\newunicodechar{·}{$\cdot$}\newunicodechar{−}{$-$}\newunicodechar{×}{$\times$}
\newunicodechar{→}{$\rightarrow$}\newunicodechar{§}{\S}\newunicodechar{≡}{$\equiv$}
\newunicodechar{≈}{$\approx$}\newunicodechar{≥}{$\geq$}\newunicodechar{≫}{$\gg$}
\newunicodechar{∈}{$\in$}\newunicodechar{⊙}{$\odot$}\newunicodechar{ℓ}{$\ell$}
\newunicodechar{µ}{$\mu$}\newunicodechar{½}{\textonehalf}\newunicodechar{†}{\dag}
\newunicodechar{…}{\ldots}\newunicodechar{¹}{\textsuperscript{1}}
\newunicodechar{²}{\textsuperscript{2}}\newunicodechar{⁰}{\textsuperscript{0}}
\newunicodechar{⁷}{\textsuperscript{7}}\newunicodechar{⁸}{\textsuperscript{8}}
\newunicodechar{⁻}{\textsuperscript{-}}\newunicodechar{₀}{\textsubscript{0}}
\newunicodechar{₁}{\textsubscript{1}}\newunicodechar{₂}{\textsubscript{2}}
\providecommand{\tightlist}{\setlength{\itemsep}{0pt}\setlength{\parskip}{0pt}}
\usepackage{dblfloatfix}

\usepackage{bookmark}
\IfFileExists{xurl.sty}{\usepackage{xurl}}{} % add URL line breaks if available
\hypersetup{
  pdftitle={FoldNTT: A Multiplier- and Twiddle-Lean NTT Core with Formally Verified Arithmetic for Proth Primes},
  pdfauthor={Masato Kamba (Nyx Foundation)},
  hidelinks,
  pdfcreator={LaTeX via pandoc}}

\title{FoldNTT: A Multiplier- and Twiddle-Lean NTT Core with Formally
Verified Arithmetic for Proth Primes}
\author{\IEEEauthorblockN{Masato Kamba}\IEEEauthorblockA{Nyx Foundation}}
\date{2026}

\begin{document}
\bstctlcite{IEEEexample:BSTcontrol}
\maketitle
\begin{abstract}
Hardware for lattice-based post-quantum cryptography spends a large
share of its area on the number-theoretic transform (NTT), dominated by
modular multipliers and twiddle storage. FoldNTT is a redesign of the
released radix-2 CFNTT accelerator (TCHES 2022) for the Falcon / FN-DSA
prime q = 12289 with one hardware multiplier per butterfly instead of
three and about half the stored twiddle constants. The Proth shape q =
3·2¹² + 1 turns modular reduction into shift-and-add K-RED folds, and
the bit-reversed twiddle table obeys
\texttt{w{[}N/2+j{]}\ =\ ψ·w{[}j{]}}, so half the table is derived
without a multiplier. Checking the released RTL against the mathematics
also exposed a bug: its inverse transform omits a per-stage halving and
returns 2¹⁰·x, which the retrofit corrects. Every arithmetic block is
proven by exact-width SMT and compositional SymbiYosys proofs,
control-safety invariants by k-induction; the proofs are mutation-tested
and the composed transform is validated by simulation, all rerun by CI.
On Artix-7 in a fully open flow, the retrofit costs 3→1 DSP48 per
butterfly and −50\% stored twiddle bits at a whole-core Fmax cost of
about 4\% (best of three seeds; within the seed-to-seed spread),
measured on the released datapath driven by a controller we
reconstructed (the reference FSM was never released) and validated by
full-core simulation; a sequential core with our own controller builds
to a timing-gated Basys-3 bitstream.
\end{abstract}

\section{1. Introduction}\label{introduction}

Lattice-based post-quantum schemes such as Kyber/ML-KEM \cite{fips203},
Dilithium/ML-DSA \cite{fips204}, and Falcon/FN-DSA \cite{falcon} spend
a large share of their cycles in polynomial multiplication, which
hardware accelerates with the number-theoretic transform (NTT), a
Fourier-style transform over integers modulo a prime q. The NTT's
hardware cost is dominated by two resources: the modular multipliers
inside the butterflies and the twiddle ROM. This paper reduces both
without changing the surrounding memory system or control.

We start from a concrete, peer-reviewed artifact: the CFNTT accelerator
\cite{cfntt}, whose contribution is a conflict-free memory mapping for
an in-place radix-2/4 NTT, released as open RTL. From it we build
FoldNTT, an NTT core that preserves the forward transform, corrects the
inverse, and uses a third of the multipliers and about half the stored
constants; every arithmetic block is proven equal to the mathematics and
the composition is validated by simulation. Three components deliver
this.

\begin{itemize}
\item
  \textbf{A verified 1-multiplier butterfly (§4.1).} For q = 12289 =
  3·2¹²+1 (a Proth prime, and Falcon's modulus), the reference's Barrett
  reduction spends two hardware multipliers beyond the unavoidable
  product. We replace them with shift-add \textbf{K-RED} folds, leaving
  one multiplier total, and fold the resulting constant factor into the
  twiddle ROM. K-RED is established in software NTTs and in hardware,
  for Kyber and for FHE-sized moduli (§2); our contribution is the
  verified drop-in retrofit, in which the same ROM fold supplies half of
  the repair for a bug we found in the released inverse transform (§3).
\item
  \textbf{A ψ-fold twiddle ROM (§4.2).} The bit-reversed negacyclic
  table obeys \texttt{w{[}N/2+j{]}\ =\ ψ·w{[}j{]}}; with ψ
  shift-friendly (Falcon ψ = 7), half the ROM is derived by a
  shift-subtract followed by a constant-threshold reduction, with no
  multiplier. This approximately halves the stored words (1023 → 512;
  the folded table keeps a scaled w{[}0{]} as a base word); the relation
  recurses to a quarter algebraically. It is distinct from the negation
  symmetry used by prior half-memory twiddle generators, which is
  unavailable here because the stored ψ-exponents span {[}0, N) while ψ
  has order 2N.
\item
  \textbf{A functional-verification methodology (§5)} that checks the
  arithmetic and datapath against a mathematical specification, down to
  the shipped ROM contents, reproducibly in CI. This differs from the
  masking/side-channel focus of recent PQC-hardware verification. Its
  abstractions are domain-faithful (the solver, not a hand argument,
  discharges the assume-guarantee seams), its control-safety proofs are
  inductive, and the proofs are mutation-tested (eight mutations over
  five harnesses). The inverse-transform bug surfaced through this
  checking, not through the shipped testbench.
\end{itemize}

The K-RED leg generalizes to other Proth NTT primes, with a generator
that emits and checks per-prime reducer RTL (§4.3), validated on Kyber
(q = 3329). §6 reports costs measured in an open FPGA flow (yosys +
openXC7 \texttt{nextpnr-xilinx}, Artix-7), with no Vivado required: at
the whole core, 3→1 DSP per butterfly and half the twiddle storage at a
Fmax cost of about 4\% (−21\% on Compact-FALCON's ENS normalized-area
metric, defined in §6), and the inverse-transform bug fixed. Because the
released radix-2 core's control FSM was never published, we
reconstructed one from the datapath's own timing and validated it by
full-core simulation (§5); the whole-core numbers are measured on that
controller. The design is one construction with two instantiations: the
streaming retrofit demonstrates the claim on the published architecture,
and a sequential single-butterfly (BFU) core with our own controller
(the ``own-FSM core'' below) packages the same verified blocks as a
self-contained accelerator that builds to a bitstream (§6).

\section{2. Background}\label{background}

\textbf{Negacyclic NTT.} Lattice schemes multiply polynomials in the
ring \texttt{R\_q\ =\ Z\_q{[}x{]}/(x\^{}N\ +\ 1)}. Naively this is an
O(N²) convolution; the NTT turns it into O(N log N) by evaluating at the
powers of a primitive N-th root of unity. The \emph{negacyclic} wrap
(the \texttt{x\^{}N\ =\ −1} quotient) is handled by pre/post-weighting
with powers of ψ, a primitive 2N-th root (\texttt{ψ²\ =\ ω},
\texttt{ψ\^{}N\ =\ −1}), so that a pointwise product in the transform
domain equals the negacyclic convolution back in \texttt{R\_q}:
\(a \cdot b = \mathrm{INTT}(\mathrm{NTT}(a) \odot \mathrm{NTT}(b))\). A
radix-2 transform is a sequence of \(\log_2 N\) stages of
\textbf{butterflies}; the forward pass uses decimation-in-time in
natural-to-bit-reversed order (DIT-NR), the inverse
decimation-in-frequency in bit-reversed-to-natural order (DIF-RN), which
lets both share one bit-reversed twiddle table and avoids an explicit
reorder: with g = 2\^{}\{9−p\} the group size at stage p and 0 ≤ k
\textless{} g, the table satisfies w{[}2g−1−k{]} = −w{[}g+k{]}⁻¹ (mod
q), so the inverse factor and its sign are both absorbed by reading the
reversed entry, which is why the inverse butterfly subtracts in the
order (v−u). The DIF-RN inverse butterfly additionally carries a
per-stage \(\tfrac12\) scaling (the \(N^{-1}\) of the inverse,
distributed one factor of \(2^{-1}\) per stage), realized by a
``multiply-by-\(2^{-1}\)'' operator \texttt{op21}, the operator the
released radix-2 core omits (§3). For Falcon/FN-DSA,
\texttt{N\ =\ 1024}, \texttt{q\ =\ 12289}, and the reference uses
\texttt{ψ\ =\ 7} (a primitive 2048-th root mod q).

\textbf{The CFNTT accelerator.} CFNTT \cite{cfntt} is an in-place,
memory-based radix-2/4 NTT accelerator whose contribution is a
\textbf{conflict-free memory mapping}: coefficients are striped across
two banks by the parity of their address (bank = XOR of address bits,
offset = address ≫ 1), and the address generator emits, for every
radix-2 stage, the pair of operands a butterfly consumes. Because the
two operands of any stage differ in exactly the stage's bit, they always
fall in different banks. Both are therefore read (and later written) in
the same cycle with no bank conflict, keeping the single pipelined
butterfly fully fed. Twiddles come from one shared ROM of
\texttt{N\ −\ 1\ =\ 1023} words in the bit-reversed layout
\texttt{w{[}i{]}\ =\ ψ\^{}\{bitrev(i)\}} (RTL address A holds
w{[}A+1{]}; w{[}0{]} = 1 is not stored), read via a small
twiddle-address generator whose sequence matches the stage/loop
counters. The released radix-2 RTL is what we retrofit; we leave its
memory system, address generators and conflict-free mapping untouched,
changing only the butterfly's arithmetic (§4.1) and the ROM's internals
(§4.2).

\textbf{Modular reduction.} Barrett and Montgomery are the
general-purpose choices. For Proth primes q = k·2\^{}m+1, K-RED
\cite{longa2016kred} reduces with shifts and adds. It is established in
software NTTs and has hardware precedent: K²-RED \cite{bisheh2021k2red}
applies it to Kyber's q = 3329, and shift-add variants (K²-RED-Shift
over Proth-ℓ primes) have been evaluated for the 32- and 64-bit moduli
of FHE \cite{tosun2024modred}.

\textbf{Twiddle storage.} Prior work reduces the ROM by on-the-fly
generation (a modular multiplier per butterfly) \cite{krieger2025tool}
or by a half-memory generator using the negation symmetry
\texttt{W\^{}\{N/2\}\ =\ −1} \cite{im2024tfg}. We use a different,
address-halving relation of the bit-reversed layout that is
multiplier-free when ψ is shift-friendly.

\textbf{Verified PQC hardware.} Recent machine-checked work on PQC
hardware targets masking composition and side-channel leakage
\cite{iskander2026a, iskander2026b}; for FHE, Casas et al.
\cite{casas2023fhe} verify a compute engine's NTT datapath and
micro-sequencer compositionally against an ISA specification (§7).

\section{3. A bug in the released inverse
transform}\label{a-bug-in-the-released-inverse-transform}

Because our methodology (§5) checks the RTL against the mathematical
transform rather than against a testbench, it surfaced a functional bug
in the released accelerator. We describe it in full, since it is the
finding that motivates the methodology.

\textbf{The defect.} The DIF-RN inverse butterfly must apply a
\(\tfrac12\) scaling per stage: the \(N^{-1}\) of the inverse transform,
distributed as one factor of \(2^{-1}\) each of the \(\log_2 N\) stages
(paper Alg. 3; the reference's own Python model applies this as
\texttt{op21}, \(x\cdot 2^{-1}\bmod q = x\,(q{+}1)/2 \bmod q\)). The
released radix-2 RTL ships \texttt{modular\_half.v} but instantiates it
nowhere in \texttt{compact\_bf.v}: in inverse mode (\texttt{sel=1}) the
butterfly computes \((u{+}v,\ (v{-}u)w)\), with w the reversed-table
entry of §2, and no halving. The radix-4 PEs (\texttt{PE0–PE3.v}) do
instantiate \texttt{modular\_half}, so the omission is specific to the
radix-2 tree.

\textbf{Consequence.} Each inverse stage is a factor of 2 too large, so
after \(\log_2 N = 10\) stages the radix-2 inverse output is scaled by
\(\mathbf{2^{10}\bmod q}\):
\(\mathrm{INTT}(\mathrm{NTT}(x)) = 2^{10}x\), not \(x\). The forward
transform is unaffected. Because the map is linear, there is no partial
cancellation; the error is exactly a global constant. The output is
therefore a valid-looking vector of residues, indistinguishable from a
correct one without a reference value.

\textbf{Bug, or an unnormalized inverse?} An inverse transform that
returns N·x and leaves N⁻¹ to be folded into a later constant is a
legitimate design choice, so the question is whether the omission is
intended. Three facts say it is not. The CFNTT paper's Alg. 3 specifies
the per-stage halving for the radix-2 inverse, and the reference's own
Python model applies it. The released radix-4 processing elements
implement it. The radix-2 tree ships \texttt{modular\_half.v} but
instantiates it in no module, and no other module of the released
radix-2 tree (\texttt{top\_poly\_mul.v} and below) applies a
compensating constant. The radix-2 RTL therefore disagrees with its own
paper and model, and we report it as a bug on that basis.

\textbf{Why testing didn't catch it.} The shipped testbench
(\texttt{tb\_top.v}) drives stimulus and reads memory files but asserts
nothing about the result, and no reference vector is committed. A single
end-to-end functional assertion would have caught the bug.

\textbf{How we found and confirmed it.} The round-trip property
\texttt{INTT(NTT(x))\ =\ x} failed in our SMT/simulation checks; the
counterexample was a clean global 2¹⁰ factor, which points directly at a
missing per-stage 2⁻¹. We confirmed it with bit-exact integer models of
the released datapath modules, each proven equivalent to its Verilog
(\texttt{verify\_radix2.py}), driven through a complete N=1024 inverse
(\texttt{bug\_intt\_halving.py} reproduces \texttt{2¹⁰·x}); localized it
to the un-instantiated \texttt{modular\_half}; and reported it upstream
(issue \#7 at \texttt{github.com/xiang-rc/cfntt\_ref}; the empty control
FSM \texttt{fsm.v} is the related issue \#4). We then reproduced it at
the full-core RTL level: the released datapath (banks, address
generators, \texttt{compact\_bf}, \texttt{tf\_ROM}), driven by the
controller we reconstructed for the missing \texttt{fsm.v} (§5), returns
\texttt{NTT(x)} exactly and \texttt{INTT(NTT(x))\ =\ 2¹⁰·x} on every
tested vector (\texttt{run\_sim.py}). Both issues are open and
unacknowledged at the time of writing.

\textbf{The fix, and what it costs here.} Reinstating the halving costs
two \texttt{modular\_half} (op21) gates per butterfly, one per output
path. In the K-RED redesign (§4.1) the multiply-path gate acts on the
twiddle word rather than on the product: the ROM stores W = 9⁻¹·w, the
inverse twiddle op21(W) = (2·9)⁻¹·w is derived from that word inside the
butterfly's existing twiddle delay chain, and the multiply then yields
((v−u)·w)/2 directly (Lemma 2). The add path gets the second gate. The
fix therefore adds no multiplier, no latency and no port change; its
cost is two shift-add gates, which a correct reference would also have
to pay. Our verified core round-trips exactly (§5, §6).

\section{4. Design}\label{design}

Both techniques are interface-compatible with the reference (same ports,
delay fabric and latencies) and are applied as a pair: the multiplier
returns 9·a·b mod q and the ROM returns 9⁻¹-scaled words, so either one
alone changes the transform. Their contracts (operands \textless{} q,
mode held constant during a transform, the corrected inverse) are stated
with the proofs in §5. Figure 1 is the proposed radix-2 butterfly; only
the shaded blocks change. We cite four small algebraic facts as Lemmas
1--4; they are stated in Appendix A, with paper proofs in the artifact
(\texttt{docs/lemmas.md}) and machine checks as the certificates.

Table 1 summarizes the design: each algebraic fact and the hardware it
saves.

\begin{table*}[!t]
\centering\footnotesize
\caption{Algebraic fact → hardware saved.}
\begin{tabular}{@{}
>{\raggedright\arraybackslash}p{(\linewidth - 4\tabcolsep) * \real{0.3333}}
  >{\raggedright\arraybackslash}p{(\linewidth - 4\tabcolsep) * \real{0.3333}}
  >{\raggedright\arraybackslash}p{(\linewidth - 4\tabcolsep) * \real{0.3333}}@{}}
\toprule\noalign{}
\begin{minipage}[b]{\linewidth}\raggedright
algebraic fact
\end{minipage} & \begin{minipage}[b]{\linewidth}\raggedright
hardware consequence
\end{minipage} & \begin{minipage}[b]{\linewidth}\raggedright
measured (§6)
\end{minipage} \\
\midrule\noalign{}
Lemma 1: K-RED fold for q = 3·2¹²+1 & 3 → 1 DSP48 per butterfly & −67\%
DSP \\
Lemma 2: halving fuses into the ROM word & inverse-transform bug fixed
with two op21 gates, no added multiplier or latency & in
\texttt{compact\_bf\_v2} \\
Lemma 3: w{[}N/2+j{]} = ψ·w{[}j{]} & twiddle ROM stores half the words,
no multiplier & −50\% stored bits; LUT 241 → 192 \\
Lemma 4: constant scalings commute & K-RED and ψ-fold compose with no
correction hardware & end-to-end exact (§5) \\
\bottomrule\noalign{}
\end{tabular}
\end{table*}

\begin{figure*}[t]
\centering
\resizebox{0.92\textwidth}{!}{%
\begin{tikzpicture}[
  >={Stealth[length=2.2mm]}, font=\small, line width=0.4pt,
  block/.style={draw, rounded corners=1pt, minimum height=8mm, inner sep=4pt, align=center},
  hi/.style   ={draw, rounded corners=1pt, minimum height=8mm, inner sep=4pt, align=center, fill=black!12},
  reg/.style  ={draw, minimum height=6.5mm, minimum width=8mm, inner sep=2pt},
  dot/.style  ={circle, fill, inner sep=1pt}]
  % ---- inputs (left), stacked u / v / w ----
  \node (u)  at (0,2.4)  {$u$};
  \node (v)  at (0,1.2)  {$v$};
  \node (w)  at (0,0)    {$W$};
  \node[reg, right=5mm of u] (du) {DFF};
  \node[reg, right=5mm of v] (dv) {DFF};
  \node[reg, right=5mm of w] (dw) {DFF};
  \node[reg, right=4mm of dw] (dw2){DFF};
  \node[block, right=6mm of dv] (mux) {mux\\[-1pt]\scriptsize(sel)};
  % ---- changed core: K-RED mult + fused half ----
  \node[hi, right=13mm of mux] (mul) {\texttt{modular\_mul}\\[-1pt]\textbf{K-RED} — \scriptsize 1 DSP (not 3)};
  \node[hi] (half) at ($(mul)+(0,-1.6)$) {\texttt{modular\_half}\\[-1pt]\scriptsize op21($W$) on ROM word};
  % ---- add / sub / op21 ----
  \node[block] (add) at ($(mul)+(5.4,0.7)$)  {\texttt{modular\_add}};
  \node[block] (sub) at ($(mul)+(5.4,-0.95)$) {\texttt{modular\_sub}};
  \node[hi, right=8mm of add] (op) {op21\\[-1pt]\scriptsize INTT $\tfrac12$};
  \node[right=8mm of op]  (bl) {\texttt{bf\_lower}};
  \node[right=13mm of sub] (bu) {\texttt{bf\_upper}};
  % ---- wires ----
  \draw[->] (u)-- (du);  \draw[->] (v)-- (dv);  \draw[->] (w)-- (dw);
  \draw[->] (dw)-- (dw2);
  \draw[->] (dv)-- (mux);
  \draw[->] (mux)-- (mul);
  \draw[->] (dw2) |- (half);
  \draw[->] (half)-- (mul);
  % product of the multiply feeds add and sub from the left
  \coordinate (pt) at ($(mul.east)+(0.5,0)$);
  \draw (mul.east) -- (pt);
  \draw[->] (pt) |- (add.west);
  \draw[->] (pt) |- (sub.west);
  \draw[->] (add)-- (op);
  \draw[->] (op)-- (bl);
  \draw[->] (sub)-- (bu);
  % u bypass: tap after its DFF, run along the TOP (clear of every block) to
  % just left of add, then drop down a rail into add and sub upper-left — u is
  % the pass-through operand of both add (u+vw) and sub (u-vw).
  \node[dot] (ud) at ($(du.east)+(0.5,0)$) {};
  \draw (du.east) -- (ud);
  \coordinate (uc)  at ($(add.west)+(-0.4,0)$);
  \coordinate (uc2) at ($(sub.west)+(-0.4,0)$);
  \draw (ud) -- (uc |- ud) -- (uc);
  \draw[->] (uc) -- (add.170);
  \draw (uc) -- (uc2);
  \draw[->] (uc2) -- (sub.170);
\end{tikzpicture}}
\caption{Proposed \texttt{compact\_bf\_v2}; shaded blocks are the changes vs the
reference. The twiddle port receives the scaled word $W = 9^{-1}w$ and the
single \textbf{K-RED} multiplier computes $9\cdot v\cdot W = v\cdot w$, replacing
the reference's three multipliers. The two \texttt{op21} ($\times\tfrac12$)
gates, one on the ROM word (\texttt{modular\_half}) and one on the INTT add
path, are the \S3 bug fix; forward mode bypasses both. Bypass muxes, delay
chains and the inverse-mode reordering (the subtraction precedes the
multiply and \texttt{bf\_upper} is the multiplier output) are not drawn.
In unscaled twiddles, \texttt{sel}=0 (NTT) yields $(u{+}vw,\;u{-}vw)$ and
\texttt{sel}=1 (INTT) yields $(\tfrac12(u{+}v),\;\tfrac12(v{-}u)\,w)$. Same
ports, delays and latency as the reference \texttt{compact\_bf}.}
\label{fig:datapath}
\end{figure*}
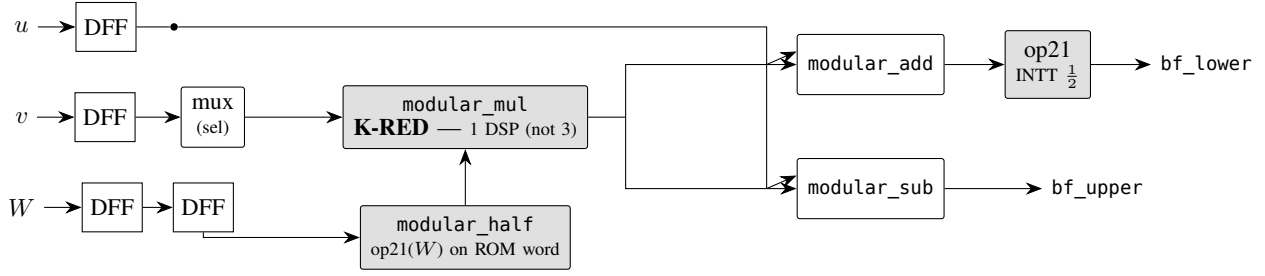

The multiplier is the single hardware multiply; the two \texttt{op21}
(modular\_half) gates, one on the ROM word and one on the add path, are
the §3 bug fix.

\subsection{4.1 K-RED butterfly}\label{k-red-butterfly}

\textbf{Reduction.} With z = z₁·2\^{}m + z₀ and k·2\^{}m ≡ −1 (mod q),
\texttt{k·z₀\ −\ z₁\ ≡\ k·z} (Lemma 1). Two folds reduce a full product
to \textless{} 2q:

\begin{verbatim}
d = 3·z[11:0] + 6q − z[27:12]   ≡ 3z,   0 < d < 2¹⁷
e = 3·d[11:0] +  q − d[16:12]   ≡ 9z,   0 < e < 2q
r = e ≥ q ? e−q : e             =  9z mod q
\end{verbatim}

\texttt{3x\ =\ (x\textless{}\textless{}1)+x}: shifts, adds, one
conditional subtraction. Latency is 4 and the ports are identical to
\texttt{modular\_mul.v}, with one hardware multiplier (the product)
instead of three.

A concrete trace, using the artifact's own constants: take v = 5555 and
the stored word W = 9⁻¹·w{[}1{]} = 3932 (w{[}1{]} = ψ\^{}bitrev(1) =
10810). The product is z = v·W = 21842260. The first fold takes z₀ =
2388, z₁ = 5332, giving d = 3·2388 + 6q − 5332 = 75566 \textless{} 2¹⁷.
The second takes d₀ = 1838, d₁ = 18, giving e = 3·1838 + q − 18 = 17785
\textless{} 2q. One subtraction finishes: r = e − q = 5496, which is 9z
mod q; because the ROM pre-scaled W by 9⁻¹, it equals v·w{[}1{]} mod q:
the ROM scaling cancels the factor.

\textbf{Absorbing the factor 9.} Each fold multiplies the residue by k,
so F folds leave a spurious factor k\^{}F; here k = 3 and F = 2, so the
factor is 9. The ROM stores W = 9⁻¹·w, so the forward butterfly's
\texttt{9·v·W\ =\ v·w} is exact; the inverse twiddle
\texttt{op21(W)=(2·9)⁻¹·w} is derived from the same word by one
\texttt{modular\_half}, and \texttt{9·(v−u)·op21(W)\ =\ ((v−u)·w)/2},
which fuses the missing halving (Lemma 2). The add path gets one more
\texttt{op21}. Pointwise multiplication (PWM; both operands are data)
double-passes the same unit with the stored constant 81⁻¹ = (k\^{}F)⁻²:
two multiplier passes per product, so PWM throughput on one unit is
halved relative to a butterfly multiply (§6 estimates the cost). The
composition is math-checked end-to-end (\texttt{kred\_math.py}:
INTT(PWM(NTT(a),NTT(b))) equals the negacyclic product), and each pass
is covered by the unit's full-domain proof; our cores do not implement
PWM in RTL, and there is no PWM-specific RTL testbench.

\subsection{4.2 ψ-fold twiddle ROM}\label{ux3c8-fold-twiddle-rom}

For the bit-reversed layout, \texttt{w{[}N/2+j{]}\ =\ ψ·w{[}j{]}} (Lemma
3), and ψ = 7 gives \texttt{7x\ =\ (x\textless{}\textless{}3)−x}. We
therefore store only the 512 lower (9⁻¹-scaled) words and derive the
upper half with a \texttt{fold7} gate, with no multiplier and the same
interface and latency as \texttt{tf\_ROM.v}:

\begin{verbatim}
t  = (base<<3) − base                       // 7·base ∈ [0, 7q)
mq = (t ≥ 6q) ? 6q : (t ≥ 5q) ? 5q : … : (t ≥ q) ? q : 0   // 6 parallel cmps
Q  = upper ? (t − mq)[13:0] : base          // one subtraction, < q
\end{verbatim}

Six parallel constant comparators pick the multiple \texttt{mq} and a
single subtraction reduces. We chose this over three chained conditional
subtractions after a logic-depth analysis (§6: LTP 31→26, area down,
still DSP-free). The relation recurses
(\texttt{w{[}N/4+j{]}\ =\ ψ²·w{[}j{]}} for 0 ≤ j \textless{} N/4), so a
quarter table is possible algebraically (checked in
\texttt{rom\_fold\_math.py}); we have not built or measured that
variant, whose derived words need factors up to ψ³ and correspondingly
more depth. The factor-9 scaling and the fold commute (Lemma 4).

§8 describes how the fold was found.

\subsection{4.3 Generalization: other Proth
primes}\label{generalization-other-proth-primes}

A generator (\texttt{kred\_gen.py}) computes, per q, the K-RED fold
count and offsets, the spurious factor k\^{}F and its inverse, and
\texttt{k·x} as shift-adds; it emits the reducer RTL and checks it. For
the ψ-fold it reports the plan (shift-add form of ψ, stored word counts)
but does not yet emit a ROM, butterfly, controller or proof harness
(§8). One caveat on scope: the K-RED leg applies to any Proth NTT prime,
but the ψ-fold's multiplier-free form additionally needs a
shift-friendly ψ (as in Falcon's ψ = 7); for a general q the fold
becomes a small constant multiply, which may not beat storing the words.
We validate Kyber, q = 3329 = 13·2⁸+1 (ML-KEM), as an independent
instance: the K-RED reducer is checked exhaustively over all z
\textless{} q², and the generated RTL passes a 60k-vector iverilog sweep
(the RTL check is a simulation sweep, not exhaustive). Kyber validates
the reducer only: ML-KEM's NTT is incomplete (q − 1 = 13·2⁸ is not
divisible by 2N = 512, so no primitive 2N-th root exists), Lemma 3's
full negacyclic table does not describe its twiddle set, and folding an
incomplete-NTT table is separate work. The generator finds a tighter
Falcon schedule than our hand-written unit, evidence that the
construction subsumes the special case.

When is the fold cheap? Two parameters decide. The fold count F stays at
2--3 whenever m is large relative to k's width, and each fold's k·x
costs adders proportional to the signed-digit weight of k. Running the
artifact's own planner (\texttt{generator/kred\_gen.py}) on primes used
in deployed schemes gives Table 2:

\begin{table}[!t]
\centering\scriptsize
\caption{K-RED fold economics for deployed NTT primes.}
\begin{tabular}{@{}
llll@{}}
\toprule\noalign{}
prime & q & folds F & k·x cost (signed-digit form) \\
\midrule\noalign{}
Falcon & 3·2¹² + 1 & 2 & 1 add \\
Kyber / ML-KEM & 13·2⁸ + 1 & 2 & 2 adds \\
Dilithium / ML-DSA & 1023·2¹³ + 1 & 2 & 1 subtraction (k = 2¹⁰−1) \\
BabyBear & 15·2²⁷ + 1 & 2 & 1 subtraction (k = \(2^4{-}1\)) \\
Goldilocks & (2³²−1)·2³² + 1 & 3 & 1 subtraction (k = 2³²−1) \\
\bottomrule\noalign{}
\end{tabular}
\end{table}

For Goldilocks, k\^{}F ≡ 1 (mod q), so no ROM scaling is needed at all.
The constant multiply k·x is therefore potentially inexpensive for NTT
primes whose k has low signed-digit weight, which covers the deployed
lattice schemes and lets an FHE or ZK deployment choose
residue-number-system (RNS) primes accordingly, but not for arbitrary
RNS primes with random large k; total cost also depends on width, fold
count and pipeline placement, which Table 2 does not capture. Table 2 is
the planner's own output (\texttt{kred\_gen.py} prints it, with a
sampled validation of each schedule). The cost column is analytical: the
emitted RTL currently uses the addition-only set-bit form of k·x, and
the signed-digit figure is the cost of the equivalent subtractive form,
not a synthesized datapath. We evaluated 14-bit and 12-bit q in RTL; the
divider-free congruence encoding is linear in the fold identities and is
expected to scale to RNS-sized moduli (30--64 bit), but we have not run
those proofs; a BabyBear (q = 15·2²⁷+1) z3 instance is the natural next
data point.

\section{5. Verification}\label{verification}

\textbf{Correctness guarantee.} The guarantee has three tiers.

\begin{itemize}
\tightlist
\item
  \emph{Proven for all inputs in scope:} each arithmetic unit over its
  full domain; the butterfly with its real delay chains, in each mode;
  the ROM at every legal address, as a 9⁻¹-scaled refinement of the
  shipped table; control-safety invariants of the own-FSM core under
  arbitrary host behaviour. Domain-faithful abstractions discharge the
  seams between these proofs.
\item
  \emph{Validated by simulation} against independent goldens: the
  composed streaming datapath (two seeded vectors), the own-FSM core
  (four round-trip and two inverse-only vectors, freshness-enforced) and
  the generated Kyber reducer (60k vectors). A monolithic proof of the
  \textasciitilde150k-cycle round trip is beyond bounded model checking
  (§8).
\item
  \emph{Outside both:} vendor-tool timing, physical hardware, and side
  channels.
\end{itemize}

We verify at three levels, all CI-reproducible, with z3 \cite{z3},
SymbiYosys \cite{yosys} and Icarus Verilog \cite{iverilog}.

\textbf{Datapath, full domain (SMT).} Exact-width z3 models of each unit
are proven equal to mod-q arithmetic over the whole input domain. The
key technique is a \textbf{divider-free congruence encoding}: instead of
asserting \texttt{r\ ==\ z\ mod\ q}, which bit-blasts a divider and
diverges past \textasciitilde24 bits, we prove the nonnegative linear
identities \texttt{3z+6q\ =\ d+z₁q}, \texttt{3d+q\ =\ e+d₁q} and
\texttt{r\ \textless{}\ q} (z = z₁·2¹² + z₀ as in §4.1, d = d₁·2¹² + d₀;
the first follows from d = 3z₀ + 6q − z₁ and 3·2¹²·z₁ = (q−1)·z₁). The
same obligation that had not converged after two hours with the divider
closes in 11 seconds.

\textbf{Pipelines, on the RTL (SymbiYosys).} The butterfly and ROM are
proven on the real Verilog with their delay chains, under the
assumptions that operands are \textless{} q and the mode input is held
constant for the proof (one harness per mode), compositionally: leaf
units are proven equivalent to behavioural models, then abstracted, so
the butterfly obligation closes in seconds. The abstractions are
\textbf{domain-faithful}: each behavioural model returns an
unconstrained value whenever an operand lies outside the range its leaf
proof justifies (\texttt{\textless{}\ q}), so the composite proof can
only pass if no leaf in the asserted cone ever sees an unreduced
operand. The assume-guarantee domain seam is thus discharged by the
solver, leaving no manual ``operands stay reduced'' argument in the
trust base; a future edit that violates it becomes a counterexample
rather than a silently unsound abstraction. Assertions are time-local:
each compares an output with the inputs \texttt{latency} cycles earlier,
gated by a saturating \texttt{guard} counter that suppresses the first
\texttt{guard} cycles after the unconstrained initial state. A bounded
model check of depth \texttt{guard+latency+1} therefore explores every
assertion window and is a complete proof; reset and single-clock/CDC are
checked structurally.

\textbf{Control, by induction (SymbiYosys).} The own-FSM core's control
plane is proven by \textbf{k-induction} with the datapath stubbed to
unconstrained sources. The invariants are the twiddle-counter closed
forms \texttt{rr\ =\ 2\^{}(9−p)\ +\ k} (forward) and
\texttt{rr\ =\ 2\^{}(10−p)\ −\ 1\ −\ k} (inverse), from which the two
external preconditions follow: every issued twiddle-ROM address lies in
the ROM's proven domain, and the two RAM write ports never target the
same address. The \texttt{busy}/\texttt{done} protocol is proven
alongside, under arbitrary host/start behaviour in both modes.
Data-independence of the control flow is structural rather than a
theorem of this proof: the FSM's next-state logic reads only its own
counters and a wait counter, and the datapath is stubbed to
unconstrained sources, so no data signal lies in the control's input
cone. The transform's cycle count is fixed by that structure and is
measured and asserted by the artifact's \texttt{run\_check.py}, which
asserts every reported count within a 70k--80k budget. This is not a
constant-time certification: the datapath is unanalyzed for power or
electromagnetic leakage (§8); the claim is only that latency and control
flow do not depend on the data. Host words are reduced mod q on load
(one conditional subtract suffices: a 14-bit word is \textless{} 2q),
and the proof asserts that every host write into the RAM is \textless{}
q. Engine writebacks are outside this proof (the datapath is stubbed);
their reducedness follows from the butterfly proof, whose outputs are
\textless{} q for inputs \textless{} q, provided the host loads all N
coefficients before starting. The proof establishes safety invariants,
not liveness: it does not show that a transform completes or that every
butterfly is issued exactly once, which the simulations cover.

\textbf{Non-vacuity.} Non-vacuity is mutation-tested: eight RTL
mutations across the five harnesses (\texttt{fv\_kred},
\texttt{fv\_bf\_v2\_ntt}, \texttt{fv\_bf\_v2\_intt},
\texttt{fv\_rom\_fold}, \texttt{fv\_core}): a fold constant, a dropped
halving gate, a skipped twiddle mux, a corrupted ROM word, a wrong fold
shift, a swapped subtraction operand, the abstraction's domain predicate
forced false (so the behavioural model returns unconstrained values for
every input), and a mis-seeded FSM counter. Each must produce a
counterexample; a harness crash does not count as a kill. This is
selected fault-detection evidence, not a proof that no obligation is
vacuous.

\textbf{System level.} The proposed modules, driven through a full
N=1024 NTT+INTT under iverilog (\texttt{run\_stream.py}, which first
reads all 1023 words out of the folded ROM and then streams them to the
butterflies, so it checks value composition rather than the ROM's live
delivery timing, which the sequential core exercises), give
\texttt{NTT(x)} = reference and \texttt{INTT(NTT(x))\ =\ x} exactly,
showing that the fix and the folded ROM values compose correctly. The
complete own-FSM core is checked more strongly (\texttt{run\_check.py}):
multi-vector round-trips (including raw 14-bit inputs ≥ q, exercising
the load reduction), the post-NTT memory compared against an
independently coded Python golden that shares only the shipped
\texttt{tf\_ROM.v} table (so a bug in the RTL arithmetic is detected on
the tested vectors), and the inverse validated by bijectivity
(\texttt{NTT\_golden(INTT\_rtl(y))\ =\ y}). That harness enforces dump
freshness: every simulation artifact is deleted before the run and
required after it, and simulator exit codes are checked. We adopted this
discipline after a repository reorganization silently disconnected an
earlier cross-check.

\textbf{Whole banked core.} The released radix-2 core cannot run as
shipped: its \texttt{fsm.v} is empty. We reconstructed a controller
(\texttt{fsm\_recon.v}) from the datapath's own timing (registered bank
and twiddle addresses at issue+1, bank and ROM reads at issue+2,
butterfly latency 6, write address and write-side select at issue+8) and
drive the complete \texttt{top\_poly\_mul}, banks and networks included,
through NTT then INTT (\texttt{run\_sim.py}, five vectors: three seeded
random, all q−1, an impulse, fresh dumps). With the shipped
\texttt{compact\_bf} and \texttt{tf\_ROM} the core returns
\texttt{NTT(x)} exactly and \texttt{INTT(NTT(x))\ =\ 2¹⁰·x} (the §3 bug
at full-core level); with \texttt{compact\_bf\_v2} and
\texttt{tf\_rom\_fold} it round-trips exactly. Both take 5290 cycles per
1024-point transform from launch to \texttt{done} (10 stages × (512
issues at one butterfly per cycle + 17 cycles of drain and turnaround)),
identical for every vector and both variants, which the harness asserts;
the host must hold the mode input until \texttt{done}, because the
shipped twiddle-address generator decodes it live. This controller is
consistent with the released datapath but is a reconstruction, not the
authors' original; its schedule, not theirs, is what the whole-core
numbers in §6 measure.

Figure 2 draws the resulting boundary; Appendix B tabulates every
obligation with its method and scope. All of it is re-run by CI on every
push.

\begin{figure}[!t]
\centering
\begin{tikzpicture}[font=\scriptsize,
  zone/.style={draw, rounded corners=1.5pt, align=left, inner sep=5pt,
               text width=0.88\columnwidth}]
\node[zone, fill=black!14] (p) {\textbf{Proven for all inputs in scope}
  (z3 + SymbiYosys)\\[1pt]
  K-RED unit, full 28-bit domain \; $\cdot$ \; fold7 \; $\cdot$ \;
  $9\cdot$ROM $\equiv$ shipped table, every legal address \; $\cdot$ \;
  butterfly, latency-exact, each mode \; $\cdot$ \;
  FSM control-safety invariants (k-induction, any host behaviour) \; $\cdot$ \;
  host writes $< q$};
\node[zone, fill=black!5, below=2mm of p] (s)
  {\textbf{Validated by simulation and mutation testing} (independent
  goldens)\\[1pt]
  composed $N{=}1024$ streaming datapath \; $\cdot$ \; own-core NTT/INTT
  round-trips \; $\cdot$ \; generated Kyber reducer \; $\cdot$ \;
  8-mutation sweep};
\node[zone, below=2mm of s] (o) {\textbf{Outside scope:}
  vendor timing \; $\cdot$ \; physical boards \; $\cdot$ \;
  power/EM side channels};
\end{tikzpicture}
\caption{The verification boundary. Everything in the top zone is proven
for every input in its stated scope; the middle zone is exercised by
simulation against independent goldens and by the mutation sweep; the
bottom zone is explicitly out of scope (\S8).}
\label{fig:boundary}
\end{figure}

The harnesses themselves are reusable: each is a self-contained
SymbiYosys or Python file parameterized per module, and the three
disciplines they encode (counterexample-only mutation kills,
domain-faithful abstraction, and BMC-complete time-local assertions) are
not specific to this core. Adapting them to another core means
re-deriving the leaf domains and control invariants; per-prime harness
generation is templated but not yet automatic (§8).

\section{6. Evaluation}\label{evaluation}

We report FPGA-primitive (\texttt{yosys\ synth\_xilinx}, 7-series) and
post-route (openXC7 \texttt{nextpnr-xilinx}, xc7a100t) numbers; the
claims rest on these. Technology-independent generic-gate counts are in
the artifact (\texttt{docs/evaluation.md}) and are quoted below only
where they differ materially from the FPGA mapping. Vendor (Vivado)
confirmation and physical on-board execution remain outside CI (§8).

Table 3 gives the measured resources, per module and for the whole core.

\begin{table*}[!t]
\centering\footnotesize
\caption{FPGA resources, per module and whole core (Artix-7, \texttt{synth\_xilinx}; LTP = longest topological path, yosys \texttt{ltp}, a logic-depth proxy).}
\begin{tabular}{@{}
>{\raggedright\arraybackslash}p{(\linewidth - 10\tabcolsep) * \real{0.1667}}
  >{\raggedright\arraybackslash}p{(\linewidth - 10\tabcolsep) * \real{0.1667}}
  >{\raggedright\arraybackslash}p{(\linewidth - 10\tabcolsep) * \real{0.1667}}
  >{\raggedright\arraybackslash}p{(\linewidth - 10\tabcolsep) * \real{0.1667}}
  >{\raggedright\arraybackslash}p{(\linewidth - 10\tabcolsep) * \real{0.1667}}
  >{\raggedright\arraybackslash}p{(\linewidth - 10\tabcolsep) * \real{0.1667}}@{}}
\toprule\noalign{}
\begin{minipage}[b]{\linewidth}\raggedright
\end{minipage} & \begin{minipage}[b]{\linewidth}\raggedright
LUT
\end{minipage} & \begin{minipage}[b]{\linewidth}\raggedright
FF
\end{minipage} & \begin{minipage}[b]{\linewidth}\raggedright
\textbf{DSP48}
\end{minipage} & \begin{minipage}[b]{\linewidth}\raggedright
RAMB18
\end{minipage} & \begin{minipage}[b]{\linewidth}\raggedright
LTP
\end{minipage} \\
\midrule\noalign{}
\texttt{modular\_mul} (Barrett) → \texttt{modular\_mul\_kred} & 29 → 83
& 101 → \textbf{74} & \textbf{3 → 1} & --- & 17 → 21 \\
\texttt{compact\_bf} (ref) → \texttt{compact\_bf\_v2} & 158 → 231 & 297
→ 270 & \textbf{3 → 1} & --- & --- \\
\texttt{tf\_ROM} → \texttt{tf\_rom\_fold} & 241 → \textbf{192} & 14 → 15
& 0 → 0 & --- & 7 → 26 \\
whole core: reference \texttt{top\_poly\_mul} & 784 & 580 & \textbf{3} &
2 & --- \\
whole core: proposed \texttt{top\_poly\_mul\_v2} & 819 & 500 &
\textbf{1} & 2 & --- \\
\bottomrule\noalign{}
\end{tabular}
\end{table*}

Figure 3 is the summary: what the retrofit changes, normalized to the
reference core. These numbers support the following observations.

\begin{figure}[!t]
\centering
\resizebox{\columnwidth}{!}{%
\begin{tikzpicture}[font=\scriptsize]
\newcommand{\perfbar}[4]{%
  \draw[fill=black!8, draw=black!30] (0,#1) rectangle (5.2,#1+0.34);
  \draw[fill=black!45, draw=black!55] (0,#1) rectangle (#2,#1+0.34);
  \node[anchor=east] at (-0.12,#1+0.17) {#3};
  \node[anchor=west] at (5.32,#1+0.17) {#4};}
\perfbar{2.55}{1.73}{DSP48 per butterfly}{33\% (3 $\to$ 1)}
\perfbar{1.85}{2.60}{stored twiddle bits}{50\%}
\perfbar{1.15}{4.12}{ENS area score}{79\% (969 $\to$ 767)}
\perfbar{0.45}{5.00}{whole-core Fmax}{96\% (143 $\to$ 138 MHz, best of 3 seeds)}
\draw[black!50, dashed] (5.2,0.3) -- (5.2,3.05)
  node[above, black, font=\scriptsize] {reference = 100\%};
\end{tikzpicture}}
\caption{The streaming retrofit vs the reference core, normalized to the
reference (100\%, dashed). Lower is better for the first three bars; Fmax
is post-route on the whole core with the reconstructed controller (\S6).
Forward transform unchanged, inverse transform corrected (\S3).}
\label{fig:summary}
\end{figure}
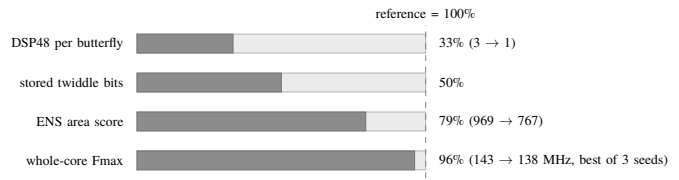

\begin{itemize}
\tightlist
\item
  The headline result is the DSP count: 3 → 1 per butterfly (−67\%) on
  real primitives, with −27\% FF on the multiplier. Each additional
  parallel butterfly costs another set of multipliers, so the saving is
  two DSP48 per instantiated butterfly; whether DSPs, bank ports,
  twiddle delivery or LUTs bound the achievable parallelism depends on
  the surrounding memory system, which we do not scale here. The
  butterfly additionally becomes inverse-correct (§3).
\item
  K-RED trades DSP for LUT/carry logic (multiplier LUTs 29 → 83; whole
  core +5\% LUT). This is favourable when DSPs bound the design and a
  small LUT cost when they do not.
\item
  On FPGA, the twiddle ROM's win is the −50\% in stored bits rather than
  a large logic cut. Under generic-gate synthesis the folded ROM is
  −79\% cells (7828 → 1611), but that figure does not transfer: at
  N=1024 the table maps to distributed LUT-ROM, where the fold saves
  ≈20\% LUT (241 → 192; fold7 adds logic). The stored-bit halving
  converts to a BRAM saving only when the halved table crosses a
  block-RAM allocation boundary; at N=1024 both 1023 and 512 words of 14
  bits fit one RAMB18.
\end{itemize}

\textbf{Timing (logic-depth proxy, \texttt{ltp}).} K-RED adds
\textasciitilde4 logic levels vs Barrett (21 vs 17, both latency-4
pipelined; Barrett's DSP hides its own multiply delay). The ψ-fold's
real cost is depth on the derived-half ROM read (LTP 26 vs 7 for a plain
lookup): a logic-depth analysis drove a redesign of \texttt{fold7} from
three chained conditional subtractions to six parallel comparators + one
subtraction (LTP 31 → 26, LUT 214 → 192, still DSP-free, re-verified).
The measured Fmax cost is small (see the post-route Fmax paragraph
below); a pipelined fold7 would remove the ROM-read depth at +1 latency.

\textbf{Whole-core area.} The last two rows of Table 3 synthesize the
entire core (one butterfly + two conflict-free banks + twiddle ROM +
address generators + FSM), reference vs proposed.

At the core level the DSP count falls 3→1 (scaling ×d with parallel
butterflies), FF falls 14\%, and LUT rises 4\% (the K-RED DSP→LUT trade
slightly exceeds the ROM's LUT saving). RAMB18 is unchanged: the two
BRAMs are the data banks, and both twiddle ROMs map to distributed
LUT-ROM, so the ψ-fold's −50\% stored bits does not cut BRAM count at
N=1024. Both cores carry the reconstructed controller that
\texttt{run\_sim.py} validates end-to-end (§5), so these are figures for
a working streaming core, with the caveat that the controller's schedule
is ours, not the CFNTT authors'.

\textbf{A complete own-FSM core, through to a bitstream.} The released
radix-2 core's control FSM is an empty file (upstream issue \#4; the
radix-4 tree does ship one), which caps any radix-2 retrofit at the
streaming/module level. We therefore also package the verified blocks
into a minimal complete accelerator with our own sequential single-BFU
FSM (§5's induction proof): one \texttt{compact\_bf\_v2}, the ψ-fold
ROM, one dual-port BRAM, and ≈74k cycles per 1024-point transform
(\textasciitilde1.5 ms at 50 MHz). This is a high-latency design point
(one butterfly at a time, no overlap between butterflies); the
conflict-free streaming schedule above is the throughput design point.
On \texttt{synth\_xilinx} it maps to 1 DSP48 + 1 RAMB18 +
\textasciitilde600 LUT / \textasciitilde186 FF, a small fraction of the
Basys-3 part (xc7a35t). The fully open flow (yosys → openXC7
\texttt{nextpnr-xilinx} → prjxray \texttt{fasm2frames} →
\texttt{xc7frames2bit}) produces a self-test bitstream whose build is
timing-gated: every clock nextpnr reports must close ≥ 50 MHz (the 100
MHz board clock is constrained; the 50 MHz core clock is a
fabric-divided clock that nextpnr reports separately), and the core
clock closes at 70--95 MHz across seeds. The self-test loads
\texttt{x{[}i{]}\ =\ 7i+1\ mod\ q}, runs NTT then INTT, and reports
\texttt{INTT(NTT(x))\ =\ x} on the LEDs. The wrapper passes in RTL
simulation; we have not run the bitstream on a board (§8).

\textbf{Post-route Fmax (open flow, no Vivado).} At the whole core the
retrofit reaches 137.5 MHz against the reference's 143.1 MHz, best of
three seeds (−4\%; \texttt{top\_poly\_mul} vs
\texttt{top\_poly\_mul\_v2}, the same RTL configurations as the area
numbers: area from a hierarchical \texttt{synth\_xilinx}, timing from a
flattened one, both with \texttt{keep} attributes on the data banks,
multipliers and ROMs, which are otherwise unobservable at the top-level
ports and would be removed). The three seeds span 127--143 MHz for the
reference and 130--138 MHz for the retrofit (medians 136 and 134 MHz),
so the difference is inside the seed-to-seed spread; each seed's nextpnr
log is archived, but three seeds do not resolve the difference and the
flow does not extract critical-path endpoints, so we read this as ``a
few percent'', not as a precise figure. That the gap is far smaller than
at the butterfly is consistent with the conflict-free memory system,
address generators and FSM, identical in both, setting the critical
path. Both netlists carry the same reconstructed controller (§5), so the
comparison is like-for-like, but the absolute figure is not CFNTT's
published one. The module-level numbers behind this, using openXC7's
\texttt{nextpnr-xilinx} + artix7 chipdb on xc7a100t, register-wrapped
modules, best of 3 seeds: \texttt{modular\_mul} (Barrett) reaches 243
MHz vs \texttt{modular\_mul\_kred} 232 MHz (−4\%), so 3→1 DSP costs
little clock speed at the multiplier; \texttt{compact\_bf} (reference)
reaches 169 MHz vs \texttt{compact\_bf\_v2} 123 MHz (−27\%), the
module-level gap that shrinks to a few percent at the core. Two effects
plausibly contribute to that gap: the reference omits the §3 halving, so
a corrected reference would also pay for those gates, and the K-RED+op21
logic lengthens the critical path vs a single DSP multiply. We have not
measured a corrected-Barrett baseline or isolated the two effects, so
this attribution is conjecture. At the module level the design trades
butterfly Fmax for DSPs and twiddle memory; adding a pipeline stage to
the K-RED path or to fold7 is an untested option that would cost one
cycle of latency and a controller change. As with area, the relative
comparisons are what the claims rest on; the absolute megahertz figures
are open-flow estimates from nextpnr-xilinx's timing model, pending
vendor static timing analysis (§8).

\textbf{Positioning vs Falcon-NTT accelerators.} CFNTT and
Compact-FALCON, the two closest designs, both target q = 12289 and both
use Barrett with full twiddle ROMs; neither of our contributions appears
in them. The two ``this work'' rows are the two instantiations of the
one construction: the streaming retrofit carries the like-for-like
comparison, the own-FSM core is the design point that executes
end-to-end. Table 4 shows the comparison.

\begin{table*}[!t]
\centering\footnotesize
\caption{Comparison with Falcon-NTT accelerators (Artix-7).}
\begin{tabular}{@{}
>{\raggedright\arraybackslash}p{(\linewidth - 12\tabcolsep) * \real{0.1429}}
  >{\raggedright\arraybackslash}p{(\linewidth - 12\tabcolsep) * \real{0.1429}}
  >{\raggedright\arraybackslash}p{(\linewidth - 12\tabcolsep) * \real{0.1429}}
  >{\raggedright\arraybackslash}p{(\linewidth - 12\tabcolsep) * \real{0.1429}}
  >{\raggedright\arraybackslash}p{(\linewidth - 12\tabcolsep) * \real{0.1429}}
  >{\raggedright\arraybackslash}p{(\linewidth - 12\tabcolsep) * \real{0.1429}}
  >{\raggedright\arraybackslash}p{(\linewidth - 12\tabcolsep) * \real{0.1429}}@{}}
\toprule\noalign{}
\begin{minipage}[b]{\linewidth}\raggedright
design
\end{minipage} & \begin{minipage}[b]{\linewidth}\raggedright
DSP
\end{minipage} & \begin{minipage}[b]{\linewidth}\raggedright
Fmax
\end{minipage} & \begin{minipage}[b]{\linewidth}\raggedright
NTT-1024 cycles / time
\end{minipage} & \begin{minipage}[b]{\linewidth}\raggedright
ENS†
\end{minipage} & \begin{minipage}[b]{\linewidth}\raggedright
formal proof
\end{minipage} & \begin{minipage}[b]{\linewidth}\raggedright
executes end-to-end
\end{minipage} \\
\midrule\noalign{}
CFNTT \cite{cfntt} (base, our flow*) & 3 & 143 MHz & 5290‡ / 37.0 µs &
969 & not reported (released inverse has the §3 defect) & RTL sim,
reconstructed FSM \\
\textbf{this work} (streaming retrofit) & \textbf{1} & 138 MHz & 5290‡ /
38.5 µs & \textbf{767} & blocks (§5) & RTL sim, reconstructed FSM \\
\textbf{this work} (own-FSM core, xc7a35t) & \textbf{1} & 70--95 MHz &
\textasciitilde74k / \textasciitilde1.5 ms @ 50 MHz & --- & blocks +
control safety (§5) & bitstream + RTL self-test (no board run) \\
Compact-FALCON \cite{dam2025compactfalcon} & 20 & 134 MHz & 640 / 4.78
µs & ≈8143 & no & as reported \\
\bottomrule\noalign{}
\end{tabular}
\par\vspace{3pt}\begin{minipage}{\linewidth}\scriptsize
*Base area and Fmax are measured on the released radix-2 datapath driven by our reconstructed controller (§5), identical in the two streaming rows; they are not CFNTT's published figures, which come from a vendor flow on the authors' own configuration and their own (unreleased) controller.\par †ENS = LUT/4 + FF/8 + BRAM×200 + DSP×100 (Compact-FALCON's own normalized area metric), computed from the area tables above. All on Artix-7; ours/base measured in the open flow (§6), Compact-FALCON as reported from Vivado. Different toolchains count LUTs differently, so the comparison that carries weight is base→ours in one flow (−21\%). Compact-FALCON is a combined FFT+NTT accelerator (17395 LUT / 7950 FF / 20 DSP / 4 BRAM), hence its far larger ENS.\par ‡Measured in RTL simulation with the reconstructed controller (§5): 10 stages × 512 butterflies at one butterfly per cycle plus per-stage drain and start/stop overhead, excluding host transfer; identical in base and retrofit and for every vector (asserted). The time column divides by the unrounded post-route Fmax (143.14 and 137.51 MHz). The own-FSM row's count is measured by \texttt{run\_check.py}.
\end{minipage}
\end{table*}

We compare against each design in turn. Against the base (same
architecture, same flow) the comparison is like-for-like: ENS −21\%
(969→767), driven by 3→1 DSP, with the forward transform unchanged, the
inverse corrected (§3), and Fmax within a few percent. One cost the
transform-level rows do not show is pointwise multiplication, which our
cores do not implement in RTL: on one K-RED unit it would pass the
multiplier twice (§4.1), so a full negacyclic product (two forward
transforms, PWM, one inverse) at N = 1024 takes an estimated 3·5290 +
2048 versus 3·5290 + 1024 cycles, about +6\%. Against Compact-FALCON we
do not claim a throughput win: it is roughly 8× faster per NTT (4.78 vs
38.5 µs), but it is a different design point, a throughput-optimized
combined FFT+NTT accelerator that is \textasciitilde10× our ENS and
spends 20 DSPs. Ours is a minimal single-BFU NTT core; its advantages
are DSP and area cost and the verification evidence, not latency. The
construction supports a parallel-BFU instantiation (independent
butterflies behind the conflict-free banks, saving two DSPs per lane),
which would trade the area lead for throughput and needs a multi-lane
conflict-free schedule and enough bank ports; we do not build or measure
it. The own-FSM core serves the opposite design point: area-constrained
deployments where a verified, self-contained core matters more than
latency.

\textbf{Energy.} The open flow provides no power model, so we make no
quantitative energy claim. Qualitatively, DSP dynamic power is a major
term in multiplier-bound NTT cores, so 3→1 DSP at equal Fmax and equal
cycle count plausibly lowers energy per transform, partially offset by
the added LUT/carry logic; vendor power estimation or board measurement
is future work (§8).

\section{7. Related work}\label{related-work}

\textbf{NTT accelerators and conflict-free memory.} In-place NTT
hardware must resolve the read/write bank conflicts of the butterfly
access pattern; CFNTT \cite{cfntt} contributes a parity-based
conflict-free mapping for radix-2/4, which we retrofit. Other
Falcon/Kyber accelerators \cite{dam2025compactfalcon, krieger2025tool}
target throughput or flexibility; the two closest Falcon-NTT designs,
CFNTT and Compact-FALCON, both use Barrett reduction with full twiddle
ROMs (§6), so neither the K-RED retrofit nor the ψ-fold appears in them.

\textbf{Modular reduction.} Montgomery and Barrett are the
general-purpose choices. K-RED \cite{longa2016kred} exploits Proth
primes \texttt{q\ =\ k·2\^{}m+1} for a shift-add reduction and is
established in software NTTs. In hardware, K²-RED
\cite{bisheh2021k2red} applies it to Kyber's q = 3329, and shift-add
variants (K²-RED-Shift over Proth-ℓ primes) have been evaluated for the
32- and 64-bit moduli of FHE \cite{tosun2024modred}. Our contribution
is a verified, drop-in retrofit of the reduction into a published
accelerator, with the residual factor folded into the twiddle ROM and
the same fold reused to correct the inverse transform.

\textbf{Twiddle storage.} Prior art shrinks the twiddle ROM by
on-the-fly generation (a modular multiplier per butterfly)
\cite{krieger2025tool} or by a half-memory generator using the negation
symmetry \texttt{W\^{}\{N/2\}\ =\ −1} \cite{im2024tfg}. The negation
symmetry is unavailable here because the table stores ψ-powers with
exponents in \texttt{{[}0,\ N)} while ψ has order 2N: no two stored
exponents differ by N. The ψ-fold instead uses the address-halving
relation \texttt{w{[}N/2+j{]}\ =\ ψ·w{[}j{]}}, which holds for a
bit-reversed power table (in natural order the corresponding factor is
ψ\^{}\{N/2\}, not shift-friendly); for a shift-friendly ψ it is
multiplier-free, it recurses to a quarter algebraically, and the
half-table RTL is proven to match the shipped ROM (up to the 9⁻¹
scaling) at every legal address.

\textbf{Verified PQC hardware.} Recent machine-checked verification of
PQC hardware targets masking and side-channel composition
\cite{iskander2026a, iskander2026b}: leakage properties rather than
functional correctness of the arithmetic against a mathematical
specification. Closest to our methodology, Casas et al.
\cite{casas2023fhe} formally verify the NTT datapath and
micro-sequencer of an FHE compute engine compositionally against an ISA
specification and report RTL bugs found in the process; their target is
a proprietary 32-bit RNS engine. We apply a similar block-plus-control
decomposition to an open, published PQC accelerator, retrofit it, and
make the contracts and every check reproducible in CI; that functional
check of a released artifact down to its ROM contents is what surfaced
the §3 bug. Our SMT/BMC/mutation toolkit uses standard techniques; the
contribution is their composition into a reproducible, whole-artifact
functional verification (proofs for the blocks and control invariants,
simulation for the composition) and its use as a design driver.

\section{8. Discussion}\label{discussion}

\textbf{How the techniques were found.} Both came out of an iterative
design loop, run as an experiment in LLM-assisted hardware design on top
of visually-3d, a tool we built that renders an architecture as a 3D
floor-plan model grounded in its RTL. One iteration of the loop: (1)
formally verify the current design; (2) regenerate the 3D model from the
verified source; (3) show rendered screenshots to a vision-language
model, which critiques the scene and looks for structure; (4) turn any
observation into a concrete design change; (5) verify the change before
accepting it. Figure 4 shows the model at five points along the loop's
37 revisions. The K-RED retrofit entered at step (4) as a conventional
optimization; the ψ-fold was noticed at step (3): once K-RED had shrunk
the arithmetic, the twiddle ROM was visibly the largest remaining block,
and the question of why half the table should not be derivable had a
mathematical answer (Lemma 3). Every candidate had to pass step (5), so
a visually suggested idea could be adopted without weakening the
correctness argument.

\begin{figure*}[t]
\centering
\includegraphics[width=\textwidth]{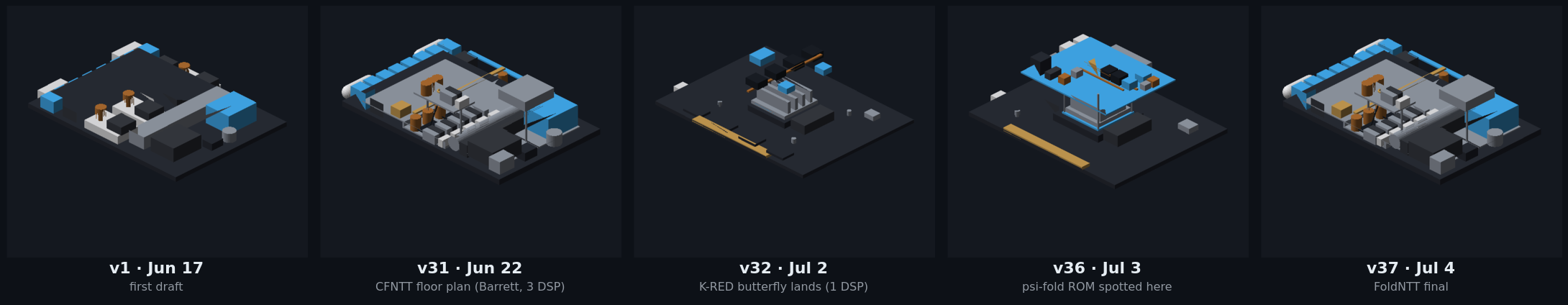}
\caption{The design loop's 3D model across its revisions: first draft
(v1); the matured CFNTT floor plan, Barrett reduction with 3 DSP per
butterfly (v31); the K-RED butterfly landing (v32); the revision in which
the ψ-fold was spotted (v36); the final FoldNTT model (v37). Each panel
is rendered by visually-3d from the RTL-grounded scene at that step.}
\label{fig:discovery}
\end{figure*}

The full model history (every revision, its renders, and the
verification verdicts between them) is published in the visually-3d
gallery at
\url{https://visually-3d.kingmasatojames.workers.dev/\#/s/ntt-fpga}, and
the rendered montage is in the repository (Figure 4), so the process is
inspectable end to end. We claim no generality for it: this is one
design, found once, with no ablation of the loop's components; the
mathematics and the proofs stand on their own.

\textbf{Limitations and future work.} Whole-core area (LUT/FF/DSP/BRAM)
and per-module post-route Fmax are both now measured via the open flow
(§6, openXC7 nextpnr-xilinx on xc7a100t). Whole-core Fmax is also
measured (143 vs 138 MHz, best of three seeds) on the core with the
reconstructed controller. The functional whole-core gap is closed by the
own-FSM core (§6): it round-trips exactly, its control-safety invariants
are proven, and it builds to a bitstream. What remains: (a) the
streaming core runs on a controller we reconstructed from the released
datapath's timing and validated by simulation (§5); it is one controller
consistent with that datapath, and the original timed behaviour of the
unreleased FSM, including its cycle count, stays unknowable, so the
whole-core figures are ours rather than CFNTT's; (b) the own-FSM core is
sequential, and a pipelined 2-bank instantiation for \textasciitilde1
butterfly/cycle is future work; (c) full-transform functional
correctness is simulation-validated, and the formal proofs cover the
listed blocks and the control-safety invariants, since a
\textasciitilde150k-cycle end-to-end BMC is out of reach; (d) physical
on-board execution and vendor (Vivado) confirmation of the open-flow
figures have not been performed; (e) power/EM side channels are out of
scope (control-flow latency is data-independent by structure, §5, but
nothing is claimed about leakage); (f) pointwise multiplication is not
implemented in RTL, and on one K-RED unit it would cost an estimated
≈6\% on a full polynomial multiplication (§6), which a second reducer or
a Barrett PWM unit would remove; and (g) the per-module Fmax flow keeps
only the best of three seeds, so timing differences of a few percent are
not resolved. Generic ψ-fold RTL emission and per-prime SymbiYosys
generation are templated but not yet automatic.

\section{9. Conclusion}\label{conclusion}

FoldNTT preserves the forward transform of the accelerator it started
from, corrects its inverse to the specification, and uses a third of the
multipliers and about half of the stored constants; every arithmetic
block is proven equal to the mathematics and the composition is
validated by simulation. Verifying the released accelerator exposed an
inverse-transform bug; the K-RED retrofit fixes it with two shift-add
gates and no added multiplier or latency, the ψ-fold halves the twiddle
ROM, and a generator extends the construction to other Proth primes. Two
instantiations demonstrate the construction: the drop-in streaming
retrofit, and a complete own-FSM core, with control-safety invariants
proven by induction, built to a timing-gated Basys-3 bitstream in a
fully open flow, whose self-test passes in RTL simulation. The proofs,
simulations and area numbers are re-run by the public repository's CI;
the Fmax and bitstream flows are one-command scripts outside hosted CI.

\section{Reproducibility}\label{reproducibility}

Everything in this paper is public at
\texttt{github.com/NyxFoundation/FoldNTT} (the retrofitted RTL, the
reference CFNTT as a submodule, all proofs, the generator, the FPGA
flow, and the rendered design-loop montage; the design-loop revision
history is at
\url{https://visually-3d.kingmasatojames.workers.dev/\#/s/ntt-fpga});
the proof, simulation and area classes are re-run by CI. All RTL,
proofs, generator and flow scripts are MIT-licensed; the upstream
\texttt{cfntt\_ref} submodule is itself MIT (© xiang-rc). A repo
\texttt{flake.nix} pins the toolchain (yosys, SymbiYosys, yices,
iverilog, uv for the z3 scripts, and the exact openXC7 tag), so
\texttt{nix\ develop} drops into a shell where every script below runs
with no further setup; hosted CI instead uses the OSS CAD Suite for the
formal flow, so the two environments differ in tool versions. Each class
of claim has a one-command reproduction; all paths below are relative to
the repository root. The GitHub Actions workflow reruns the proof,
simulation and area classes on every push. The Fmax flow runs under the
flake; the bitstream script additionally resolves yosys and the prjxray
converters through the nix registry rather than the flake lock. Both
stay outside hosted CI (the place-and-route chip database is too large
for hosted runners), as do the board-wrapper simulation and the own-core
synthesis (\texttt{ntt-core/README.md}):

\begin{itemize}
\tightlist
\item
  \textbf{Functional verification} (\texttt{run\_all.sh}): the
  exact-width z3 proofs (K-RED unit, fold7), the SymbiYosys proofs
  (butterfly miter with domain-faithful abstractions, scaled ROM
  equivalence, own-FSM control safety by k-induction, reset/CDC, the
  8-mutation sweep), the iverilog streaming round-trip
  (\texttt{verification/fullcore/run\_stream.py}), the own-core checks
  (\texttt{ntt-core/run\_check.py}) and the generator
  (\texttt{generator/kred\_gen.py}, \texttt{generator/gen\_check.py});
  the reference-datapath equivalence suite and the §3 bug reproduction
  are \texttt{verification/reference-fv/run\_all.sh} and
  \texttt{verification/bug\_intt\_halving.py}.
\item
  \textbf{Own-FSM core} (§5, §6): \texttt{ntt-core/run\_check.py} runs
  the freshness-enforced multi-vector round-trip, independent-golden
  NTT/INTT checks, and the streaming cross-validation;
  \texttt{ntt-core/fv\_core.sby} runs the control-safety induction
  proof.
\item
  \textbf{Area} (§6): \texttt{fpga/fpga\_cost.sh} (per module) and
  \texttt{fpga/fpga\_cost\_core.sh} (whole core), via
  \texttt{yosys\ synth\_xilinx}.
\item
  \textbf{Post-route Fmax} (§6): \texttt{fpga/fmax.sh} and
  \texttt{fpga/fmax\_core.sh}, via openXC7 \texttt{nextpnr-xilinx} on
  Artix-7 \texttt{xc7a100t}; no Vivado, no vendor download. The
  \texttt{flake.nix} pins the working openXC7 tag
  (\texttt{github:openXC7/toolchain-nix/0.8.2}) and exports
  \texttt{NP}/\texttt{CHIPDB}, so under \texttt{nix\ develop} both
  scripts run argument-free.
\item
  \textbf{Bitstream} (§6): \texttt{ntt-core/bit.sh} runs the full
  Vivado-free Basys-3 flow, timing-gated (every reported clock ≥ 50
  MHz).
\end{itemize}

Table 5 lists the deliverables.

\begin{table*}[!t]
\centering\footnotesize
\caption{Deliverables and their certificates.}
\begin{tabular}{@{}
>{\raggedright\arraybackslash}p{(\linewidth - 6\tabcolsep) * \real{0.2500}}
  >{\raggedright\arraybackslash}p{(\linewidth - 6\tabcolsep) * \real{0.2500}}
  >{\raggedright\arraybackslash}p{(\linewidth - 6\tabcolsep) * \real{0.2500}}
  >{\raggedright\arraybackslash}p{(\linewidth - 6\tabcolsep) * \real{0.2500}}@{}}
\toprule\noalign{}
\begin{minipage}[b]{\linewidth}\raggedright
deliverable
\end{minipage} & \begin{minipage}[b]{\linewidth}\raggedright
file
\end{minipage} & \begin{minipage}[b]{\linewidth}\raggedright
contract
\end{minipage} & \begin{minipage}[b]{\linewidth}\raggedright
certified by
\end{minipage} \\
\midrule\noalign{}
K-RED multiplier & \texttt{kred-butterfly/modular\_mul\_kred.v} & ports
of \texttt{modular\_mul.v}, latency 4 & \texttt{verify\_kred.py} (z3,
full domain), \texttt{fv\_kred.sby} \\
verified butterfly & \texttt{kred-butterfly/compact\_bf\_v2.v} &
ports/latency (6) of \texttt{compact\_bf.v} &
\texttt{fv\_bf\_v2\_\{ntt,intt\}.sby} \\
ψ-fold twiddle ROM & \texttt{psi-fold-rom/tf\_rom\_fold.v} & interface
of \texttt{tf\_ROM.v} & \texttt{fv\_rom\_fold.sby} (every address) \\
complete core & \texttt{ntt-core/ntt\_core.v} & host load/read,
start/done & \texttt{fv\_core.sby}, \texttt{run\_check.py} \\
per-prime generator & \texttt{generator/kred\_gen.py} & emits reducer
RTL + checks & exhaustive (Kyber reducer) \\
verification harnesses & \texttt{verification/}, \texttt{*/fv\_*.sby} &
per-module, self-contained & 8-mutation sweep \\
\bottomrule\noalign{}
\end{tabular}
\end{table*}

A \texttt{Dockerfile} provides a separate verification-only image
(proofs and simulations, pinned to a nixpkgs revision rather than the
flake lock, without the openXC7 flow). The numbers in this paper were
taken from the repository at commit \texttt{2571d66} (reference
submodule \texttt{8373a66}); a Zenodo DOI will be minted from the tagged
release. The single source for this paper (\texttt{docs/paper/paper.md})
builds to the canonical two-column IEEEtran PDF (\texttt{make} in
\texttt{docs/paper/}) and to a single-column draft
(\texttt{make\ draft.pdf}).

\section{Appendix A: the four lemmas}\label{appendix-a-the-four-lemmas}

Paper proofs are short and live with the artifact
(\texttt{docs/lemmas.md}); the machine checks are the certificates.
Throughout, q = k·2\^{}m + 1 is a Proth prime (k odd, k \textless{}
2\^{}m), N = 2\^{}n with 2N \textbar{} q − 1, ψ a primitive 2N-th root
of unity mod q, and \texttt{bitrev\_n} the n-bit reversal; equalities
between residues are in Z\_q.

\textbf{Lemma 1 (K-RED fold).} For all \(z \ge 0\) with
\(z = z_1 2^m + z_0\), \(0 \le z_0 < 2^m\):
\texttt{k·z₀\ −\ z₁\ ≡\ k·z\ (mod\ q)}. Iterating F folds, with
multiple-of-q offsets keeping every term nonnegative, yields r ≡
k\^{}F·z; F and the offsets are chosen per prime so that r \textless{}
2q (two folds for q = 12289 from z \textless{} 2²⁸;
\texttt{kred\_gen.py} computes the bound recurrence), and one
conditional subtraction finishes. \emph{Machine check:}
\texttt{verify\_kred.py} (z3, full 28-bit domain);
\texttt{generator/kred\_gen.py} (Kyber, exhaustive z \textless{} q²).

\textbf{Lemma 2 (INTT-halving fusion).} If the ROM stores W =
(k\^{}F)⁻¹·w, the forward multiply k\^{}F·(v·W) = v·w is exact, and
feeding op21(W) = (2k\^{}F)⁻¹·w to the inverse butterfly gives
k\^{}F·((v−u)·op21(W)) = ((v−u)·w)/2, the per-stage \(\tfrac12\) the
DIF-RN inverse requires, from the same multiply. \emph{Machine check:}
\texttt{fv\_bf\_v2\_intt.sby}, \texttt{run\_stream.py}.

\textbf{Lemma 3 (ψ-fold).} For the bit-reversed layout
\(w[i] = \psi^{\mathrm{bitrev}_n(i)}\) and \(0 \le j < N/2\):
\(w[N/2+j] = \psi \cdot w[j]\), recursing on sub-halves
(\(w[N/4+j] = \psi^2 \cdot w[j]\) for \(0 \le j < N/4\), \ldots). The
negation symmetry \(\psi^{e+N} = -\psi^e\) of prior half-memory
generators does not apply: the stored exponents lie in {[}0, N) while ψ
has order 2N, so no two differ by N. For shift-friendly ψ the derived
half is multiplier-free (ψ = 7:
\texttt{7x\ =\ (x\textless{}\textless{}3)\ −\ x}). \emph{Machine check:}
\texttt{rom\_fold\_math.py}, \texttt{verify\_rom\_fold.py},
\texttt{fv\_rom\_fold.sby} (9·RTL ≡ shipped ROM mod q at every legal
address A \textless{} 1023).

\textbf{Lemma 4 (composition).} Constant scalings mod q commute, so the
(k\^{}F)⁻¹ ROM scaling and the ψ-fold coexist:
\(\psi \cdot ((k^F)^{-1} w[j]) = (k^F)^{-1} w[N/2+j]\), and the
k\^{}F-scaling butterfly restores exactly the specified transform.
\emph{Machine check:} \texttt{rom\_fold\_math.py},
\texttt{run\_stream.py} (end-to-end).

\section{Appendix B: verification
obligations}\label{appendix-b-verification-obligations}

The detail behind Figure 2; every row is re-run by CI on each push.

\begin{table*}[!t]
\centering\footnotesize
\caption{Verification obligations, methods and scopes.}
\begin{tabular}{@{}
>{\raggedright\arraybackslash}p{(\linewidth - 4\tabcolsep) * \real{0.3333}}
  >{\raggedright\arraybackslash}p{(\linewidth - 4\tabcolsep) * \real{0.3333}}
  >{\raggedright\arraybackslash}p{(\linewidth - 4\tabcolsep) * \real{0.3333}}@{}}
\toprule\noalign{}
\begin{minipage}[b]{\linewidth}\raggedright
property
\end{minipage} & \begin{minipage}[b]{\linewidth}\raggedright
method
\end{minipage} & \begin{minipage}[b]{\linewidth}\raggedright
scope
\end{minipage} \\
\midrule\noalign{}
K-RED unit == k\^{}F·a·b mod q & z3, divider-free congruence & full
28-bit domain \\
fold7 == 7·x mod q & z3, congruence & full domain (x\textless q) \\
9·\texttt{tf\_rom\_fold} ≡ shipped \texttt{tf\_ROM} (mod q) & SymbiYosys
miter & after every enabled read of a legal address (A \textless{}
1023); outputs hold while REN is low \\
butterfly (NTT / INTT) == spec & SbY compositional, domain-faithful &
u,v,w \textless{} q, mode fixed per harness, after the flush window;
latency-exact \\
own-FSM control safety (§5) & SymbiYosys k-induction, datapath stubbed &
arbitrary host behaviour, both modes; safety only \\
host writes reduced mod q & k-induction assert, h\_din unconstrained &
every host write, symbolic data \\
reset / power-up-X / single-clock & SymbiYosys + netlist audit &
structural \\
non-vacuity & 8 RTL mutations & each kills its proof \\
streaming datapath INTT(NTT(x))=x & iverilog simulation & two seeded
vectors, N=1024 \\
own core NTT / INTT vs independent golden & iverilog + golden,
bijectivity & 4 round-trip + 2 inverse vectors incl.~raw ≥ q, fresh
dumps \\
generalization (Kyber q=3329 reducer) & exhaustive + iverilog & all
z\textless q², generated RTL (60k vectors) \\
\bottomrule\noalign{}
\end{tabular}
\end{table*}

\nocite{yosys, iverilog, z3}
\bibliographystyle{IEEEtran}
\bibliography{references}

@IEEEtranBSTCTL{IEEEexample:BSTcontrol,
  CTLdash_repeated_names = "no"
}

@article{cfntt,
  title={CFNTT: Scalable Radix-2/4 NTT Multiplication Architecture with an Efficient Conflict-free Memory Mapping Scheme}, 
  volume={2022}, 
  url={https://tches.iacr.org/index.php/TCHES/article/view/9291}, 
  DOI={10.46586/tches.v2022.i1.94-126}, 
  abstractNote={Number theoretic transform (NTT) is widely utilized to speed up polynomial multiplication, which is the critical computation bottleneck in a lot of cryptographic algorithms like lattice-based post-quantum cryptography (PQC) and homomorphic encryption (HE). One of the tendency for NTT hardware architecture is to support diverse security parameters and meet resource constraints on different computing platforms. Thus flexibility and Area-Time Product (ATP) become two crucial metrics in NTT hardware design. The flexibility of NTT in terms of different vector sizes and moduli can be obtained directly. Whereas the varying strides in memory access of in-place NTT render the design for different radix and number of parallel butterfly units a tough problem. This paper proposes an efficient conflict-free memory mapping scheme that supports the configuration for both multiple parallel butterfly units and arbitrary radix of NTT. Compared to other approaches, this scheme owns broader applicability and facilitates the parallelization of non-radix-2 NTT hardware design. Based on this scheme, we propose a scalable radix-2 and radix-4 NTT multiplication architecture by algorithm-hardware co-design. A dedicated schedule method is leveraged to reduce the number of modular additions/subtractions and modular multiplications in radix-4 butterfly unit by 20% and 33%, respectively. To avoid the bit-reversed cost and save memory footprint in arbitrary radix NTT/INTT, we put forward a general method by rearranging the loop structure and reusing the twiddle factors. The hardware-level optimization is achieved by excavating the symmetric operators in radix-4 butterfly unit, which saves almost 50% hardware resources compared to a straightforward implementation. Through experimental results and theoretical analysis, we point out that the radix-4 NTT with the same number of parallel butterfly units outperforms the radix-2 NTT in terms of area-time performance in the interleaved memory system. This advantage is enlarged when increasing the number of parallel butterfly units. For example, when processing 1024 14-bit points NTT with 8 parallel butterfly units, the ATP of LUT/FF/DSP/BRAM n radix-4 NTT core is approximately 2.2 × /1.2 × /1.1 × /1.9 × less than that of the radix-2 NTT core on a similar FPGA platform.}, 
  number={1}, 
  journal={IACR Transactions on Cryptographic Hardware and Embedded Systems}, 
  author={Chen, Xiangren and Yang, Bohan and Yin, Shouyi and Wei, Shaojun and Liu, Leibo}, 
  year={2021}, 
  month={Nov.}, 
  pages={94–126} 
}

@inproceedings{longa2016kred,
  author = {Longa, Patrick and Naehrig, Michael},
  title = {Speeding up the Number Theoretic Transform for Faster Ideal Lattice-Based Cryptography},
  year = {2016},
  isbn = {978-3-319-48964-3},
  publisher = {Springer-Verlag},
  address = {Berlin, Heidelberg},
  url = {https://doi.org/10.1007/978-3-319-48965-0_8},
  doi = {10.1007/978-3-319-48965-0_8},
  booktitle = {Cryptology and Network Security: 15th International Conference, CANS 2016, Milan, Italy, November 14-16, 2016, Proceedings },
  pages = {124–139},
  numpages = {16},
  location = {Milan, Italy}
}

@inproceedings{bisheh2021k2red,
  author={Bisheh-Niasar, Mojtaba and Azarderakhsh, Reza and Mozaffari-Kermani, Mehran},
  booktitle={2021 IEEE 28th Symposium on Computer Arithmetic (ARITH)}, 
  title={High-Speed NTT-based Polynomial Multiplication Accelerator for Post-Quantum Cryptography}, 
  year={2021},
  volume={},
  number={},
  pages={94-101},
  doi={10.1109/ARITH51176.2021.00028}
}

@misc{tosun2024modred,
  author = {Tosun, Tolun and K{\i}rb{\i}y{\i}k, Selim and Ko{\c c}er, Emre and Alaybeyo{\u g}lu, Ersin},
  title = {Optimized FPGA Architecture for Modular Reduction in NTT},
  year = {2025},
  isbn = {978-3-032-15540-5},
  publisher = {Springer-Verlag},
  address = {Berlin, Heidelberg},
  url = {https://doi.org/10.1007/978-3-032-15541-2_7},
  doi = {10.1007/978-3-032-15541-2_7},
  booktitle = {Lightweight Cryptography for Security and Privacy: 6th International Workshop, LightSec 2025, Istanbul, T{\"u}rkiye, September 1–2, 2025, Revised Selected Papers},
  pages = {117–137},
  numpages = {21},
  location = {Istanbul, T{\"u}rkiye}
}

@article{im2024tfg,
  author  = {Nari Im and Heehun Yang and Yujin Eom and Seong-Cheon Park and Hoyoung Yoo},
  title   = {Efficient Twiddle Factor Generators for {NTT}},
  journal = {Electronics},
  volume  = {13},
  number  = {16},
  pages   = {3128},
  year    = {2024},
  doi     = {10.3390/electronics13163128}
}

@inproceedings{dam2025compactfalcon,
  author={Dam, Duc-Thuan and Tran, Thai-Ha and Nguyen, Trong-Hung and Hoang, Trong-Thuc and Pham, Cong-Kha},
  booktitle={2025 IEEE International Symposium on Circuits and Systems (ISCAS)}, 
  title={Compact FALCON FFT/NTT Accelerator for Post-Quantum Cryptography}, 
  year={2025},
  volume={},
  number={},
  pages={1-5},
  doi={10.1109/ISCAS56072.2025.11043460}
}

@article{krieger2025tool,
  author={Krieger, Florian and Hirner, Florian and Can Mert, Ahmet and Sinha Roy, Sujoy},
  journal={IEEE Transactions on Computer-Aided Design of Integrated Circuits and Systems}, 
  title={A Flexible Hardware Design Tool for Fast Fourier and Number-Theoretic Transformation Architectures}, 
  year={2026},
  volume={45},
  number={5},
  pages={2531-2543},
  doi={10.1109/TCAD.2025.3595834}
}

@inproceedings{casas2023fhe,
  author={Casas, Jeremy and Yang, Zhenkun and Wang, Wen and Yang, Jin and Godbole, Adwait},
  booktitle={2023 60th ACM/IEEE Design Automation Conference (DAC)}, 
  title={Towards A Formally Verified Fully Homomorphic Encryption Compute Engine*}, 
  year={2023},
  volume={},
  number={},
  pages={1-6},
  doi={10.1109/DAC56929.2023.10247836}
}

@misc{iskander2026a,
  author        = {Ray Iskander and Khaled Kirah},
  title         = {Structural Dependency Analysis for Masked {NTT} Hardware: Scalable Pre-Silicon Verification of Post-Quantum Cryptographic Accelerators},
  howpublished  = {arXiv:2604.15249},
  year          = {2026}
}

@misc{iskander2026b,
  author        = {Ray Iskander and Khaled Kirah},
  title         = {Fresh Masking Makes {NTT} Pipelines Composable: Machine-Checked Proofs for Arithmetic Masking in {PQC} Hardware},
  howpublished  = {arXiv:2604.20793},
  year          = {2026}
}

@misc{fips203,
  author       = {{National Institute of Standards and Technology}},
  title        = {Module-Lattice-Based Key-Encapsulation Mechanism Standard ({ML-KEM})},
  howpublished = {FIPS 203},
  year         = {2024},
  doi          = {10.6028/NIST.FIPS.203}
}

@misc{fips204,
  author       = {{National Institute of Standards and Technology}},
  title        = {Module-Lattice-Based Digital Signature Standard ({ML-DSA})},
  howpublished = {FIPS 204},
  year         = {2024},
  doi          = {10.6028/NIST.FIPS.204}
}

@misc{falcon,
  author       = {Pierre-Alain Fouque and Jeffrey Hoffstein and Paul Kirchner and Vadim Lyubashevsky and Thomas Pornin and Thomas Prest and Thomas Ricosset and Gregor Seiler and William Whyte and Zhenfei Zhang},
  title        = {{Falcon}: Fast-Fourier Lattice-based Compact Signatures over {NTRU}},
  howpublished = {Specification v1.2, NIST Post-Quantum Cryptography Standardization, Round 3},
  year         = {2020},
  note         = {\url{https://falcon-sign.info/}}
}

@misc{yosys,
  author       = {Claire Wolf},
  title        = {{Yosys} Open {SYnthesis} Suite and {SymbiYosys}},
  howpublished = {\url{https://github.com/YosysHQ/yosys}, \url{https://github.com/YosysHQ/sby}},
  year         = {2024}
}

@misc{iverilog,
  author       = {Stephen Williams},
  title        = {{Icarus Verilog}},
  howpublished = {\url{https://github.com/steveicarus/iverilog}},
  year         = {2024}
}

@inproceedings{z3,
  author    = {Leonardo de Moura and Nikolaj Bj{\o}rner},
  title     = {{Z3}: An Efficient {SMT} Solver},
  booktitle = {Tools and Algorithms for the Construction and Analysis of Systems (TACAS 2008)},
  series    = {LNCS},
  volume    = {4963},
  pages     = {337--340},
  year      = {2008},
  doi       = {10.1007/978-3-540-78800-6_24}
}
\end{document}